\documentclass[12pt]{amsart}
\usepackage[T1]{fontenc}
\usepackage[centertags]{amsmath}
\usepackage{amsfonts}
\usepackage{amssymb}
\usepackage{amsthm}
\usepackage{newlfont}
\usepackage{epsfig}
\usepackage{amscd}

\newcommand{\DD}{{\mathbb D}}

\newcommand{\CC}{{\mathbb C}}

\newcommand{\ZZ}{{\mathbb Z}}
\newcommand{\KK}{{\mathbb K}}
\newcommand{\FF}{{\mathbb F}}

\newcommand{\WW}{{\mathbb W}}

\newcommand{\calC}{{\mathcal C}}

\newtheorem{Th}{Theorem}
\newtheorem{Cor}[Th]{Corollary}

\newtheorem{Prop}[Th]{Proposition}
\theoremstyle{definition}
\newtheorem{Def}{Definition}
\theoremstyle{remark}
\newtheorem*{Rem}{Remark}
\begin{document}

\title[The discrete KP equation over a finite field]{Algebro-geometric 
solution\\ of the discrete KP equation over a 
finite field\\ out of a hyperelliptic curve}
\author[M. Bia{\l}ecki]{Mariusz Bia{\l}ecki}
\address{M. Bia{\l}ecki\\ 
Instytut Geofizyki PAN\\ ul. Ksi\c{e}cia Janusza 64\\ 01-452 Warszawa\\ 
Poland}
\email{bialecki@igf.edu.pl} 
\author[A. Doliwa]{Adam Doliwa}
\address{A. Doliwa\\ Wydzia{\l} Matematyki i Informatyki \\ Uniwersytet
Warmi\'n\-sko--Mazurski\\ ul. \.Zo{\l}nierska 14A\\ 10-561 Olsztyn\\ Poland}
\email{doliwa@matman.uwm.edu.pl}
\subjclass[2000]{14H70, 37K10, 37B15}
\keywords{Integrable systems; cellular automata; finite fields;
algebraic curves; discrete KP equation}

\begin{abstract}

\noindent We transfer the algebro-geometric method of construction of
solutions of the discrete KP equation to the finite field case.
We emphasize role of the Jacobian of the underlying algebraic curve 
in construction of the solutions.
We illustrate in detail the procedure on example of a hyperelliptic curve.

\bigskip

\end{abstract}

\maketitle

\section{Introduction} 
Cellular automata are dynamical systems on a lattice with values being discrete
(usually finite) as well. They are one of the more popular and distinctive
classes of models of complex systems. Introduced in various  contexts \cite{Ulam-CA,vNeumann-CA} around 1950
 they have found wide applications in 
different areas, from physics through
chemistry and biology to social sciences \cite{Wolfram}. 

One of the most interesting properties of cellular automata is that complex
patterns can emerge from very simple uptade rules. However, usually one cannot 
easily predict how a given cellular automaton
will behave without going through a number of time steps on a computer.
Due to their completely
discrete nature, cellular automata are naturally suitable for computer
simulations, but also here it would be instructive to have examples of rules
with large classes of analytical solutions, integrals of motions and
other "integrable features".

The problem of construction of integrable cellular automata is not new and
was undertaken in a
number of papers (see, for example, 
\cite{Fokas-aut,BBGP,BruschiSantiniCA,TS-aut}). In particular in 
\cite{TTMS,MSTTT} it was given
a systematic method, called ultra-discretization, of obtainig
cellular automaton version of a given discrete integrable system. 

Recently a new
approach to integrable cellular automata was proposed in
\cite{DBK}. Its main idea is to keep the form of a given
integrable discrete system but to transfer the algebro-geometric method of 
construction of its solutions~\cite{Krichever-acnde,BBEIM}
from the complex field $\CC$ to a finite field. This method, 
which in principle
can be applied to any integrable discrete system with known algebro-geometric
method of solution, has been
applied to the fully discrete 2D Toda system (the Hirota equation) and in
\cite{BialDol-d-KP-KdV} to discrete KP and KdV equations (in Hirota form). In
particular,
the finite field valued multisoliton solutions of these equations have been
constructed. 
We remark that the algebraic geometry over finite fields, although
conceptually similar 
to that over the field of complex numbers~\cite{Hartshorne}, has its 
own tools and
peculiarities~\cite{ArithGeom,Moreno}.
It is also nowadays very
important in practical use in modern approaches to public
key cryptography \cite{Koblitz} and in the theory of 
error correcting codes \cite{Stichtenoth}.

The aim of this paper is to study in the finite field context the very distinguished 
example of integrable system --- the
discrete KP equation. We present the algebro-geometric scheme of construction
of its solutions in a finite field and we demonstrate
its linearization on the level of the abstract Jacobi
variety of the corresponding algebraic curve. We 
illustrate details of the construction on example of a hyperelliptic curve. 

We remark that in \cite{BBGP} it was observed that the Lax
representation of the discrete
sine-Gordon equation of Hirota \cite{HirotaSG} has a meaning also when the field
of complex numbers is replaced by a finite field of the form $\FF_{p^2}$, where
$p$ is a prime number. The authors of \cite{BBGP} showed also that, in principle, 
the corresponding integrals of motion can be calculated. Finally, we note that  
possibility of considering the soliton theory in positive characteristic has
been anticipated in \cite{Mumford}.

The organization of the paper is as follows. In Section \ref{sec:ffKP-alg-geom} we 
first summarize the finite field version of Krichever's construction of
solutions of the discrete KP equation, then we present its abstract Jacobian
picture. In Section \ref{sec:hyp}
we apply the method 
starting from an algebraic curve of genus two.

\section{The finite field solution of the discrete KP equation out of nonsingular
algebraic curves} \label{sec:ffKP-alg-geom}
We first briefly recall algebro-geometric construction of solutions
of the discrete KP equation over finite fields \cite{DBK,BialDol-d-KP-KdV}. We 
discuss in addition a possible degeneracy of the linear problem and its
consequences. Then we present the Jacobian picture of the 
construction, in which the integrable nature of the equation is evident. 
We point out some aspects of the representation which will help us to construct
effectively solutions of the equation.

\subsection{General construction} \label{sec:gen-constr}
Consider an algebraic projective curve $\calC/\KK$ (or simply $\calC$),  
absolutely irreducible, nonsingular, of genus $g$, defined over the finite  
field $\KK=\FF_q$ with $q$ elements, where $q$ is a power of a 
prime integer $p$ (see, for example \cite{Stichtenoth,Hartshorne}). 
By $\calC(\KK)$ we denote 
the set of $\KK$-rational points of the curve. 
By $\overline{\KK}$ denote the algebraic closure of  
$\KK$, i.e., $\overline{\KK} = \bigcup_{\ell=1}^\infty \FF_{q^\ell}$, and by 
$\calC(\overline{\KK})$ denote the corresponding infinite set (often identified
with $\calC$) of 
$\overline{\KK}$-rational points of the curve.  
The action of the Galois group $G(\overline{\KK}/\KK)$ (of automorphisms of  
$\overline{\KK}$ which are identity on $\KK$, see \cite{Lang-alg}) extends 
naturally to action on $\calC(\overline{\KK})$. 
 
Let us choose:\\ 
1. four points $A_i\in\calC(\KK)$, \  $i=0,1,2,3$, \\ 
2. effective $\KK$-rational  divisor of order $g$,
i.e., $g$ points $B_\gamma\in\calC(\overline{\KK})$, $\gamma=1,\dots,g$, 
which satisfy the following $\KK$-rationality condition 
\[ 
\forall \sigma\in  G(\overline{\KK}/\KK), \quad  
\sigma(B_\gamma) = B_{\gamma^\prime}.
\] 
As a rule we assume here that all the 
points used 
in the construction are distinct and in general position. In particular, 
the divisor $\sum_{\gamma=1}^g B_\gamma$ is non-special. 
\begin{Def}  \label{def:psi}
Fix $\KK$-rational local parameter $t_0$ at $A_0$. 
For any integers $n_1,n_2,n_3\in \ZZ$ define the function  
$\psi(n_1,n_2,n_3)$ as a 
rational function on the curve $\calC$ with the following properties:\\ 
1. it has pole of the order at most $n_1+n_2+n_3$ at $A_0$,\\  
2. the first nontrivial coefficient of its expansion in $t_0$ at $A_0$ is 
normalized to one, \\  
3. it has zeros of order at least $n_i$ at $A_i$ for $i=1,2,3$, \\  
4. it has at most simple poles at points $B_\gamma$, $\gamma=1,\dots,g$.
\end{Def} 
As usual, zero (pole) of a negative order means pole (zero) of the 
corresponding positive order. Correspondingly one should exchange the 
expressions "at most" and "at least" in front of the orders of poles and 
zeros. 
By the standard (see e.g., \cite{BBEIM}) application of the Riemann--Roch 
theorem (and the general position
assumption) we conclude that  
the wave function $\psi(n_1,n_2,n_3)$ exists and is unique. 
The function $\psi(n_1,n_2,n_3)$ is $\KK$-rational, which follows  
from $\KK$-rationality conditions of sets of points in their definition. 
\begin{Rem}
In what follows we will often normalize functions in a sense of point~2 of
Definition~\ref{def:psi}. 
\end{Rem}
Fix $\KK$-rational local parameters $t_i$ at $A_i$, $i=1,2,3$. 
In the generic case, which we assume in the sequel, when the order of  
the pole of $\psi(n_1,n_2,n_3)$ at $A_0$ is 
$n_1+n_2+n_3$ denote by $\zeta^{(i)}_k(n_1,n_2,n_3)$, $i=0,1,2,3$, 
the $\KK$-rational
coefficients of expansion of $\psi(n_1,n_2,n_3)$ at $A_i$, respectively, 
i.e.,
\begin{align*}
\psi(n_1,n_2,n_3) &= \frac{1}{t_0^{(n_1+n_2+n_3)}} 
\left( 1 + \sum_{k=1}^{\infty} \zeta^{(0)}_k(n_1,n_2,n_3) 
t_0^k \right), \\ 
\psi (n_1,n_2,n_3) &= t_i^{n_i} \sum_{k=0}^{\infty} 
\zeta^{(i)}_k(n_1,n_2,n_3)  t_i^k , \quad i=1,2,3.  
\end{align*} 
Denote by $T_i$ the operator of translation 
in the variable $n_i$, $i=1,2,3$, 
for example $T_1 \psi(n_1,n_2,n_3) = \psi(n_1+1,n_2,n_3) $. 
Uniqueness of the wave function implies the following statement. 
\begin{Prop} \label{prop:equations-psi} 
Generically, the function $\psi$  
satisfies equations 
\begin{equation}  
T_i \psi - T_j \psi +  
\frac{T_j\zeta^{(i)}_ 0}{\zeta^{(i)}_0} \psi = 0 , \quad i\ne j , 
\quad i,j=1,2,3. 
\label{eq:psi}  
\end{equation} 
\end{Prop} 
\begin{Rem}
When the genericity assumption fails then the linear problem~\eqref{eq:psi} 
degenerates, i.e., some of its terms are absent. 
\end{Rem}
Notice that equation \eqref{eq:psi} gives
\begin{equation} \label{eq:zeta-zeta}
\frac{T_j\zeta^{(i)}_ 0}{\zeta^{(i)}_0} = -  
\frac{T_i\zeta^{(j)}_ 0}{\zeta^{(j)}_0} , \quad i\ne j,\quad i,j=1,2,3.  
\end{equation}
Define 
\begin{equation} \label{eq:rho-def}
\rho_i= (-1)^{\sum_{j<i} n_j} \zeta^{(i)}_0, \quad i=1,2,3,
\end{equation} 
then equation
\eqref{eq:zeta-zeta} implies existence of a $\KK$-valued potential 
(the $\tau$-function) defined (up to a multiplicative
constant) by formulas 
\begin{equation} \label{eq:tau-def} 
 \frac{T_i\tau}{\tau} = \rho_i,\quad  i=1,2,3.  
\end{equation} 
Finally, equations \eqref{eq:psi} give rise to condition 
\begin{equation}  \label{eq:KP-rho}
\frac{T_2\rho_1}{\rho_1} - \frac{T_3\rho_1}{\rho_1} + 
\frac{T_3\rho_2}{\rho_2} = 0, 
\end{equation}  
which written in terms of the $\tau$-function gives the discrete KP
equation~\cite{Hirota} called also the Hirota equation
\begin{equation}  \label{eq:tauKP} 
(T_1\tau) \;(T_2T_3\tau) - (T_2\tau) \;(T_3T_1 \tau) + 
(T_3\tau) \;(T_1T_2\tau) = 0. 
\end{equation} 
\begin{Cor} \label{cor:KP-exp}
Equation \eqref{eq:KP-rho} can be obtained also from expansion of equation 
\eqref{eq:psi} at $A_k$, where $k=1,2,3$, $k\ne i,j$.
\end{Cor}
\begin{Rem}
Absence of a term in the linear problem \eqref{eq:psi} 
(see the remark after Proposition \ref{prop:equations-psi}) reflects, due to
Corollary~\ref{cor:KP-exp}, in absence of
the corresponding term in equation \eqref{eq:tauKP}. This implies that in the
non-generic case, when we have not defined the $\tau$-function yet, we are 
forced to put it to zero. Let us notice in advance (see the next section)
that it is in complete
analogy with the well known (complex field)
interpretation of the algebro-geometric solution $\tau$ of the discrete 
KP equation as, essentially, the Riemann 
theta function.
\end{Rem}
\subsection{The Jacobian interpretation}
Denote by $\mathrm{Div}(\calC)$ the abelian group of the
divisors on the curve $\calC$
and by $\mathrm{Pic}^0(\calC)$ the group of eqivalence classes of
degree zero divisors $\mathrm{Div}^0(\calC)$ modulo the
principal divisors. There exists~\cite{Lang,Milne} an abelian variety
$J(\calC)$ of dimension $g$ (the Jacobian of the curve) and an 
injective map
$\phi:\calC\to J(\calC)$ (the Abel map) such that the extension of 
$\phi$ to
$\mathrm{Div}(\calC)$ establishes an isomorphism between 
$\mathrm{Pic}^0(\calC)$
and $J(\calC)$. Moreover, if there exists a $\KK$-rational point 
$A\in\calC(\KK)$ of the curve,
then $\phi$ can be defined by
\[ \calC\ni P\mapsto \phi(P)=[P-A]\in J(\calC),
\]
where $[P-A]$ designates the class of the degree zero divisor $P-A$ in
$\mathrm{Pic}^0(\calC)$.

Denote by $\DD_r(\calC)$ the effective 
divisors of degree $r$ of the curve
$\calC$ and by $\phi_r$ the extension of $\phi$ to $\DD_r(\calC)$
\[ \DD_r(\calC) \ni D\mapsto \phi_r(D)=[D-r\cdot A]\in J(\calC).
\]
The direct image of $\phi_r$ is a subvariety $\WW_r$ of dimension $r$ if 
$0\leq r
\leq g$, and of dimension $g$ if $r>g$. In particular, $\WW_{g-1}$ defines a
divisor in $J(\calC)$.

The group $\mathrm{Pic}^0(\calC;\KK)$ of eqivalence classes of
$\KK$-rational degree zero divisors $\mathrm{Div}^0(\calC;\KK)$ modulo the
principal $\KK$-rational divisors can be identified with the abelian group 
$J(\calC;\KK)$ of $\KK$-rational points of the Jacobian variety. For finite
field $\KK$ the group $\mathrm{Pic}^0(\calC;\KK)$ is finite as well and its
order can be found using properties of the zeta function of the
curve (see, for example \cite{Stichtenoth}).

Let us present in this picture the description of the wave function $\psi$ and
of the $\tau$-function. We choose the point $A_0$ as the reference
point $A$ and consider the following divisor 
$D(n_1,n_2,n_3)\in\mathrm{Div}^0(\calC;\KK)$ of degree zero
\[ D(n_1,n_2,n_3)=n_1(A_0-A_1)+ n_2(A_0-A_2)+ n_3(A_0-A_3) +
\sum_{\gamma=1}^g B_\gamma - g\cdot A_0,
\]
with linear dependence on $n_1$, $n_2$ and $n_3$.
Its equivalence class in $\mathrm{Pic}^0(\calC;\KK)$ has 
the unique $\KK$-rational representant
of the form
\[ X(n_1,n_2,n_3) = \sum_{\gamma=1}^g X_\gamma(n_1,n_2,n_3) - g\cdot A_0 .
\] 
This equivalence is given by a function 
whose divisor  reads
\begin{equation} \label{eq:div-psi}
n_1(A_1-A_0)+ n_2(A_2-A_0)+ n_3(A_3-A_0) + 
\sum_{\gamma=1}^g X_\gamma (n_1,n_2,n_3) -
\sum_{\gamma=1}^g B_\gamma \sim 0 .
\end{equation}
If we normalize such a function at $A_0$ according to
Definition~\ref{def:psi} it becomes the wave function $\psi$. 
Notice that some of $X_\gamma$ could be $A_0$ which would mean that
$[D(n_1,n_2,n_3)]\in \WW_{g-1}$.
This correspondence gives rise to a set of
important facts.
\begin{Cor} Evolutions in variables $n_1$, $n_2$ and $n_3$
define linear flows in the Jacobian.  
\end{Cor}
\begin{Cor}
Points $X_\gamma$, $\gamma=1,...,g$ indicate 
zeros of
the wave function which are not specified in the previous construction. 
\end{Cor}
\begin{Cor}
If
$[D(n_1,n_2,n_3)]\in \WW_{g-1}$ then the pole of the wave function 
at $A_0$ has the order less then
$(n_1+n_2+n_3)$, i.e., we are in the non-generic case, thus 
$\tau(n_1,n_2,n_3)=0$.
\end{Cor}
\begin{Rem}
Notice that because $[D(0,0,0)]\not\in \WW_{g-1}$ then $\tau(0,0,0)\not= 0$.
\end{Rem}
\begin{Rem}
If $\KK=\CC$ then the algebraic curve $\calC$ is the compact Riemann
surface, theorems of Abel and Jacobi identify the Jacobian with quotient of
$\CC^g$ by the period lattice, and a theorem of Riemann identifies 
$\WW_{g-1}$ with a certain translate of the zero
locus of the Riemann theta function (see~\cite{GrHa}). Then, 
as we mentioned in remark after Corollary~\ref{cor:KP-exp}, the 
algebro-geometric solution $\tau$ of the discrete KP equation 
becomes, with appropriate understanding of its argument via 
the divisor $D(n_1,n_2,n_3)$ and up to a not essential and
non-vanishing multiplier, the Riemann theta function
(see, e.g.~\cite{KWZ}).
In particular, in such an interpretation the zeros of 
$\tau$ are
located in points of the translate $\WW_{g-1}$ of the Theta divisor.
\end{Rem}

Let us discuss periodicity of solutions of the finite
field version of the KP equation
obtained using the above method. Denote by $\Pi_i$,
$i=1,2,3$, the ranks of cyclic subgroups of $J(\calC;\KK)$ generated by divisors
$A_i - A_0$, then for arbitrary $k_i\in\ZZ$,
$i=1,2,3$, 
\begin{equation*}
D(n_1+k_1\Pi_1, n_2+k_2\Pi_2, n_3+k_3\Pi_3) \sim D(n_1, n_2, n_3).
\end{equation*}
In particular, $\tau(n_1, n_2, n_3)=0$ implies 
$\tau(n_1+k_1\Pi_1, n_2+k_2\Pi_2, n_3+k_3\Pi_3)=0$.

There exist unique (normalized at $A_0$) functions $h_i$, $i=1,2,3$, 
with zeros of order $\Pi_i$ at $A_i$,
poles of order $\Pi_i$ at $A_0$ and no other singularities and zeros such that
\begin{equation} \label{eq:q-p-psi}
\psi(n_1+k_1\Pi_1, n_2+k_2\Pi_2, n_3+k_3\Pi_3) = 
h_1^{k_1} h_2^{k_2}h_3^{k_3}\,\psi(n_1,n_2,n_3). 
\end{equation}
\begin{Rem}
Generalizing above considerations, if for $\ell_i\in\ZZ$, $i=1,2,3$, 
\begin{equation*}
\sum_{i=1}^3 \ell_i(A_i - A_0) \sim 0,
\end{equation*}
then $(\ell_1,\ell_2,\ell_3)$ is the period vector of zeros of the
$\tau$-function and vector of quasi-periodicity (in the above sense)
of the wave function.
\end{Rem}
Equation \eqref{eq:q-p-psi} implies quasi-periodicity of the 
functions $\zeta_0^{(i)}$, $i=1,2,3$,
\begin{equation*}
\zeta_0^{(i)}(n_1+k_1\Pi_1, n_2+k_2\Pi_2, n_3+k_3\Pi_3) = 
c_{(i)1}^{k_1} c_{(i)2}^{k_2}c_{(i)3}^{k_3}\,\zeta_0^{(i)}(n_1,n_2,n_3),
\end{equation*}
with the (non-zero) factors $c_{(i)j}\in\KK_*$ equal to
\begin{equation*}
c_{(i)j} = \left(\frac{h_j}{t_j^{\delta_{ij}\Pi_j}}\right)\Bigr\rvert_{P=A_i}.
\end{equation*}
The multiplicative group $\KK_*$ is a cyclic group of order $q-1$, therefore
the functions $\zeta_0^{(i)}$ are periodic. Their periods in variable $n_j$ 
are equal to $\Pi_j$ times the order of the subgroup of $\KK_*$ generated by
$c_{(i)j}$ (a divisor of $q-1$).
Due to possible change of sign (see equation \eqref{eq:rho-def})
the periods of $\rho_i$
can be eventually doubled with respect to the corresponding periods of
$\zeta_0^{(i)}$. Again, periodicity of $\rho_i$ implies quasi-periodicity of
$\tau$ with a factor from $\KK_*$, thus the period of $\tau$ in variable $n_i$ 
can be maximally $q-1$ times the period of $\rho_i$ in that variable. 

\section{A "hyperelliptic" solution of the discrete KP equation}
\label{sec:hyp}
Our goal here is to demonstrate how does the method described above work.
We perform all steps of the construction (see also \cite{Bial-phd} for details)
starting from a given algebraic
curve, which we have chosen to be a hyperelliptic curve, 
due to relatively simple description of Jacobians of such
curves~\cite{MWZ-hyp}. We consider a hyperelliptic curve of genus $g=2$
but the technical tools used here can be applied directly to 
hyperelliptic curves of arbitrary genus. 

\subsection{A hyperelliptic curve and its Jacobian}
Consider a hyperelliptic curve $\calC$ of genus $g=2$  
defined over the field $ \FF_7 $ and given by the equation
\begin{equation}\label{eq:curve} 
\calC: \quad v^2 + u v = u^5 + 5 u^4 + 6 u^2 + u + 3.
\end{equation}
The $(u,v)$ coordinates of its  $\FF_7$-rational points are presented in 
Table~\ref{tab:PunktyF7}. The curve has one point at infinity, denoted by
$\infty$, whose preimage on the nonsingular model of $\calC$ consists of one
point only \cite{Sha}, and where the local uniformizing parameter can be chosen 
as  $u^2/v$ ($u$ is a polynomial function of order $2$, and $v$ is a polynomial
function of order $5$).
The point opposite (with respect to the hyperelliptic
authomorphism) to $P$ is denoted by $\tilde{P}$. 
The only two special point of the curve are
$(6,4)$ and the infinity point $\infty$. 

\begin{table}
\begin{center}
\begin{tabular}{|c||c|c|}
\hline \hline
$i$ &  $ P_i$  & $\tilde P_i$  \\
\hline \hline
0 &   $\infty$   & $P_0 $\\
1 &   $(1,1)$    & $(1,5)$    \\
2 &   $(2,2)$    & $(2,3)$    \\
3 &   $(5,3)$    & $(5,6)$    \\
4 &   $(6,4)$    & $P_4$ \\
\hline \hline
\end{tabular}
\bigskip
\caption{$\FF_7$-rational points of the curve $\calC$.}
\label{tab:PunktyF7}
\end{center}
\end{table}
We identify the field $\FF_{49}$ as the extension of $\FF_7$ by the polynomial
$x^2+2$, i.e., $\FF_{49}={\FF_7[x]}/(x^2+2)$. 
Let us introduce the following notation: the element $k \in \FF_{49}$ 
represented by the polynomial $ \beta x + \alpha $ is
denoted by the {\em natural} number $7 \beta + \alpha $. 
The Galois group $G(\FF_{49}/\FF_7)=\{ \mathrm{id}, \sigma  \}$, where 
$\sigma$ is the Frobenius automorphism, acts  
on elements of $\FF_{49}\setminus\FF_7$ in the following way:
\[ \FF_{49}\setminus\FF_7\ni k = 7 \beta +\alpha \mapsto 
\sigma(k) = 7(7-\beta )+\alpha .
\]
The coordinates of
$\FF_{49}$-rational points of the curve (which are not $\FF_7$-rational) 
are presented in Table~\ref{tab:PunktyF49}. 

\begin{table}
\begin{center}
\begin{tabular}{|c||c|c|c|c|}
\hline \hline
$i$ &   $P_i$ & $\tilde P_i$ & $  P_i^\sigma  $ &  $ \tilde P_i^\sigma  $  \\
\hline \hline
5 &   $(0,21)$ & $(0,28)$ & $ \tilde P_5 $ & $ P_5$ \\
6 &    $(3,9)$ & $(3,44)$ & $\tilde P_6$ & $ P_6$ \\
7 &   $(4,26)$ & $(4,33)$ & $\tilde P_7 $ & $P_7$ \\
\hline
8 &   $(7,5)$ & $(7,44)$ & $(42,5)$ & $(42,9)$ \\
9 &   $(8,22)$ & $(8,26)$ & $(43,29)$ & $(43,33)$ \\
10 &   $(11,5)$ & $(11,47)$ & $(46,5)$ & $(46,12)$ \\
11 &  $(12,6)$ & $(12,45)$ & $(47,6)$ & $(47,10)$  \\
12 &   $(13,14)$ & $(13,29)$ & $(48,35)$ & $(48,22)$ \\
13 &   $(14,8)$ & $(14,34)$ & $(35,43)$ & $(35,27)$ \\
14 &   $(15,13)$ & $(15,28)$ & $(36,48)$ & $(36,21)$ \\
15 &   $(16,17)$ & $(16,23)$ & $(37,38)$ & $(37,30)$ \\
16 &   $(17,0)$ & $(17,39)$ & $(38,0)$ & $(38,18)$ \\
17 &   $(18,4)$ & $(18,41)$ & $(39,4)$ & $(39,20)$ \\
18 &   $(19,9)$ & $(19,28)$ & $(40,44)$ & $(40,21)$ \\
19 &   $(20,12)$ & $(20,31)$ & $(41,47)$ & $(41,24)$ \\
20 &   $(22,4)$ & $(22,30)$ & $(29,4)$ & $(29,23)$ \\
21 &   $(25,6)$ & $(25,32)$ & $(32,6)$ & $(32,25)$ \\
22 &   $(27,7)$ & $(27,22)$ & $(34,42)$ & $(34,29)$ \\
\hline \hline
\end{tabular}
\bigskip
\caption{$\FF_{49}$-rational points of the curve $\calC$ 
(which are not $\FF_7$-rational); here
$\tilde P $ is the opposite of $P$, and $P^\sigma$ denotes its conjugate with
respect to the lift 
of the Frobenius automorphism.}
\label{tab:PunktyF49}
\end{center}
\end{table}

In the next step we find the group of the $\FF_7$-rational points 
$J(\calC;\FF_7)$ of the Jacobian of the 
curve. The number of its points can be found from the number of 
$\FF_7$-rational
and $\FF_{49}$-rational points of the curve by application of properties of the
zeta function of the curve $\calC$ (see for instance 
\cite{Stichtenoth,Koblitz}). In
our case the curve has $8$ $\FF_7$-rational points and $74$ $\FF_{49}$-rational
points which implies the following form of the zeta function 
$\zeta(\calC;T)$
\[  \zeta(\calC;T)=\frac{P(T)}{(1-T)(1-7T)}, \qquad 
P(T)=1+12 T^2+49 T^4 .
\]
The number $ \# J(\calC;\FF_7)$ of the $\FF_7$-rational points points of the 
Jacobian is equal to  $P(1)=62$,
and therefore $J(\calC;\FF_7)$ is the direct sum of cyclic groups of orders 
$31$ and $2$. 

Let us choose the infinity point $\infty$ as the basepoint. 
The group law in the Jacobian of a hyperelliptic curve 
can be intuitively described in a way which is a
higher-genus analog of the well known addition operation for points of elliptic
curves.   
We present here only its sketch for genus $g=2$ and in the generic case
of addition of two points of $J(\calC)$ with representants of the form
\begin{equation*}
E_i = Q_i + R_i - 2\infty, \quad i=1,2, 
\end{equation*}
with all points distinct. If $E_3 = Q_3 + R_3 - 2\infty$ is the representant of
$[E_1 + E_2]$, i.e., 
\begin{equation*}
E_1 + E_2 = (g) + E_3,
\end{equation*}
then 
\begin{equation} \label{eq:div-D1D2D3}
E_1 + E_2 + \tilde{E}_3 \sim 0,
\end{equation}
where we have used the fact that for any point $P\in\calC(\overline\KK)$ of a
hyperelliptic curve
the divisor $P + \tilde{P} - 2\infty$ is principal. Therefore, there exists a
normalized polynomial function $f$ of the order six, thus necessarily of the form
\begin{equation*}
f = a+ b u + c u^2 + d u^3 + e v,
\end{equation*}
with divisor given by the left hand side of equation
\eqref{eq:div-D1D2D3}.
Its zeros at $Q_i$ and $R_i$, $i=1,2$, and the normalization condition
allow to fix the coefficients and then to find two other zeros $\tilde{Q}_3$ 
and $\tilde{R}_3$. 

Geometrically, we are looking for two other intersection points
of the cubic interpolating known four points with the hyperelliptic curve. In
cases when some points of $E_1+E_2$ are repeated, the interpolation step must 
be
adjusted to ensure tangency to the curve with sufficient multiplicity. When
divisors have less points then we consider the interpolating curve of lower
degree (some intersection points are at infinity).
Finally, the transition function $g$ is the unique normalized function with
the nominator equal to $f$ and the
denominator being the normalized polynomial function with
the divisor $E_3+\tilde{E}_3$.

The full
description of the group $J(\calC;\FF_7)$
is given in Table~\ref{tab:jakobianN}. The
divisor $D_1=P_1 -\infty$ generates the subgroup of order $31$ and the
divisor $D_4=P_4 -\infty$ generates the subgroup of order $2$.
We present the reduced representants $[nD_1 + mD_4]_r$ of elements $[nD_1 + mD_4]$ of 
$J(\calC;\FF_7)$, where $n\in\{ 0,1,\dots,30 \}$ and $ m\in\{ 0,1 \}$.
Moreover we write down the transition functions $g_m(n)$ defined by the 
following divisor equation
\begin{equation} \label{eq:g}
[nD_1 + mD_4]_r+D_1=(g_m(n)) + [(n+1)D_1 + mD_4]_r. 
\end{equation}
Also the transition functions for the sums $[nD_1 +mD_4]_r+D_4$ 
can be read off from Table~\ref{tab:jakobianN}.
In particular, to find such a transition function (we call it $W(0,1)$)
with $n=29$ and $m=0$, i.e.,
\begin{equation} \label{eq:W01-1}
(1,5) + (1,5) + D_4 - 2\infty = (W(0,1)) + (12,6) + (47,6) - 2\infty,
\end{equation}
we make use of the
fact that the analogous transition function for $n=30$ and $m=0$ is $1$.
Then
\begin{equation} \label{eq:W01-2}
W(0,1) = g_0(29) \cdot 1 \cdot [g_1(29)]^{-1} = \frac{2+3u+4u^2+v}{6+4u+u^2},
\end{equation} 
where in the last equality we have used equation \eqref{eq:curve} of the curve 
to get rid of
$v$ from the denominator.
\begin{table}
\begin{center}
\begin{tabular}{|r||r|c||r|c|} 
 \hline \hline 
$n$  & \hspace{1cm} $[n D_1]_r$ \hspace{1cm}   &
$g_0(n)$ & \hspace{.5cm} $[n D_1 + D_4 ]_r$ \hspace{1cm} & $g_1(n)$ \\
\hline \hline
0 &    $  0 \;$             & 1
	&$ (6,4) - \;\infty\:  $  & 1\\
1 &  $  \boldsymbol{(1,1)-  \;\infty}\: $     &  1
	&$ (1,1)+(6,4)- 2 \infty $ & $ \frac{5+5 u+3  {u^2}+v}{6+4  u+{u^2}} $	\\
2 & $ (1,1)+(1,1)- 2 \infty $  &  $\frac{u+5  {u^2}+v}{{{(2+u)}^2}} $  
	& $ (12,45)+(47,10)- 2 \infty $  &  $\frac{1+5  {u^2}+v}{2+5  u+{u^2}} $  \\
3 & $ (5,6)+(5,6)- 2 \infty $   &  $\frac{1+u+4  {u^2}+v}{(2+u) (5+u)} $ 
	& $ (15,28)+(36,21)- 2 \infty  $   & $\frac{6  {u^2}+v}{2+{u^2}} $ \\
4 & $ (2,3)+(5,3)- 2 \infty $   &  $\frac{2+4  {u^2}+v}{5+4  u+{u^2}} $ 
	& $ (7,44)+(42,9)- 2 \infty $   & $ \frac{5+u+v}{4+6  u+{u^2}} $  \\
5 & $ (19,9)+(40,44)- 2 \infty $   &   $\frac{4 u+2  {u^2}+v}{5+5  u+{u^2}} $
	& $ (11,5)+(46,5)- 2 \infty $   & $\frac{6+6 u+{u^2}+v}{3+6  u+{u^2}} $ \\
6 & $ (22,4)+(29,4)- 2 \infty $   &  $\frac{5+2  u+6  {u^2}+v}{(2+u)  (5+u)} $ 
	& $ (18,41)+(39,20)- 2 \infty $   & $\frac{5+3  u+5  {u^2}+v}{5+3  u+{u^2}} $  \\
7 & $ (2,3)+(5,6)- 2 \infty $      & $\frac{5+6 u+2  {u^2}+v}{5+2  u+{u^2}} $
	& $ (16,17)+(37,38)- 2 \infty $   & $\frac{5+4 u+4  {u^2}+v}{3+u+{u^2}} $  \\
8 & $ (27,22)+(34,29)- 2 \infty $      &$\frac{1+3  u+2  {u^2}+v}{1+{u^2}} $
	& $ (17,39)+(38,18)- 2 \infty $   & $\frac{3+2  u+{u^2}+v}{(5+u)  (6+u)} $   \\
9 & $ (14,34)+(35,27)- 2 \infty $      &$\frac{1+5 u+v}{(1+u)  (5+u)} $
	& $ (1,5)+(2,2)- 2 \infty $   &  $6+u $ \\
10 & $ (2,2)+(6,4)- 2 \infty $     &$\frac{3+5  u+5  {u^2}+v}{{{(5+u)}^2}} $
	& $ \boldsymbol{(2,2)-  \;\infty} \:$   & $1$ \\
11 & $ (2,3)+(2,3)- 2 \infty $     & $\frac{6+u+6 {u^2}+v}{3+2  u+{u^2}} $
	& $ (1,1)+(2,2)- 2 \infty $  &  $\frac{4+2  {u^2}+v}{3+5  u+{u^2}} $  \\
12 & $ (13,14)+(48,35)- 2 \infty $     & $\frac{3+6  u+4  {u^2}+v}{2+2  u+{u^2}} $
	& $ (8,22)+(43,29)- 2 \infty $    & $\frac{2+4  u+v}{(2+u) (6+u)} $\\
13 & $ (20,12)+(41,47)- 2 \infty $     & $\frac{5 u+{u^2}+v}{(1+u)  (2+u)} $
	& $ (1,5)+(5,3)- 2 \infty $   & $6+u $   \\
14 & $ (5,3)+(6,4)- 2 \infty $     &$ \frac{6+5  u+2  {u^2}+v}{6+6  u+{u^2}} $ 
	& $ \boldsymbol{(5,3) - \;\infty} \:$   & $1$   \\
15 & $ (25,32)+(32,25)- 2 \infty $     &$\frac{5+{u^2}+v}{6+6 u+{u^2}} $
	& $ (1,1)+(5,3)- 2 \infty $  &  $\frac{u+5  {u^2}+v}{(2+u) (6+u)} $  \\
16 & $ (25,6)+(32,6)- 2 \infty $     &$\frac{6+5  u+2  {u^2}+v}{(1+u)  (2+u)} $
	& $ (1,5)+(5,6)- 2 \infty $  & $6+u $    \\
17 & $ (5,6)+(6,4)- 2 \infty $     &$\frac{5 u+{u^2}+v}{2+2  u+{u^2}} $
	& $ (5,6) -  \;\infty \:$   &  $1$  \\ 
18 & $ (20,31)+(41,24)- 2 \infty $     &$\frac{3+6  u+4  {u^2}+v}{3+2  u+{u^2}} $ 
	& $ (1,1)+(5,6)- 2 \infty $   &  $\frac{2+4  u+v}{3+5 u+{u^2}} $ \\
19 & $ (13,29)+(48,22)- 2 \infty $     &$\frac{6+u+6 {u^2}+v}{{{(5+u)}^2}} $
	& $ (8,26)+(43,33)- 2 \infty $   &  $\frac{4+2  {u^2}+v}{(5+u)  (6+u)} $\\
20 & $ (2,2)+(2,2)- 2 \infty $     &$\frac{3+5  u+5  {u^2}+v}{(1+u)  (5+u)} $ 
	& $ (1,5)+(2,3)- 2 \infty $  &   $6+u $ \\
21 & $ (2,3)+(6,4)- 2 \infty $     &$\frac{1+5 u+v}{1+{u^2}} $
	& $ (2,3)- \;\infty\: $    & $ 1 $\\
22 & $ (14,8)+(35,43)- 2 \infty $     &$\frac{1+3  u+2  {u^2}+v}{5+2  u+{u^2}} $
	& $ (1,1)+(2,3)- 2 \infty $   & $\frac{3+2  u+{u^2}+v}{3+u+{u^2}}$  \\
23 & $ (27,7)+(34,42)- 2 \infty $     &$\frac{5+6 u+2  {u^2}+v}{(2+u)  (5+u)} $
	& $ (17,0)+(38,0)- 2 \infty $   &  $\frac{5+4 u+4  {u^2}+v}{5+3  u+{u^2}} $\\
24 & $ (2,2)+(5,3)- 2 \infty $     &$\frac{5+2  u+6  {u^2}+v}{5+5  u+{u^2}} $
	& $ (16,23)+(37,30)- 2 \infty $  &  $\frac{5+3  u+5  {u^2}+v}{3+6  u+{u^2}}$  \\
25 & $ (22,30)+(29,23)- 2 \infty $     &$\frac{4 u+2  {u^2}+v}{5+4  u+{u^2}} $ 
	& $ (18,4)+(39,4)- 2 \infty $  &   $\frac{6+6  u+{u^2}+v}{4+6  u+{u^2}}$\\
26 & $ (19,28)+(40,21)- 2 \infty $     &$\frac{2+4  {u^2}+v}{(2+u)  (5+u)} $
	& $ (11,47)+(46,12)- 2 \infty $  &  $\frac{5+u+v}{2+{u^2}}$  \\
27 & $ (2,2)+(5,6)- 2 \infty $     &$\frac{1+u+4 {u^2}+v}{{{(2+u)}^2}}$ 
	& $ (7,5)+(42,5)- 2 \infty $   & $\frac{6  {u^2}+v}{2+5 u+{u^2}} $ \\
28 & $ (5,3)+(5,3)- 2 \infty $     &$\frac{u+5u^2+v}{(6+u)^2}$
	& $ (15,13)+(36,48)- 2 \infty $   & $\frac{1+5  {u^2}+v}{6+4  u+{u^2}}$ \\
29 & $ (1,5)+(1,5)- 2 \infty $     & $(6+u)$  
	& $ \boldsymbol{(12,6)+(47,6)- 2 \infty} $  &   $\frac{5+5  u+3 {u^2}+v}{(1+u)  (6+u)}  $   \\
30 & $ (1,5)-  \;\infty \:$     &$(6+u)$
	& $ (1,5)+(6,4)- 2 \infty $   & $6+u $  \\
\hline \hline

\end{tabular}
\bigskip
\caption{The group $J(\calC;\FF_7)$ of $\FF_7$-rational points of the 
Jacobian  as the simple sum of its cyclic subgroups.}
\label{tab:jakobianN}
\end{center}
\end{table}

\subsection{Construction of the wave and $\tau$ functions}

In order to find a solution of the discrete KP equation let us fix the
following points of the curve $\calC$,  
\begin{equation*} 
A_0=\infty, \quad A_1=(1,1), \quad A_2=(2,2), \quad A_3=(5,3),
\end{equation*}  
with the uniformizing parameters
\begin{equation*} 
t_0=u^2/v, \quad t_1=u-1, \quad t_2=u-2, \quad t_3=u-5,
\end{equation*}  
and
\begin{equation*}
B_1=(12,6), \quad B_2=(47,6).
\end{equation*}
Because
\begin{gather*}
A_0 - A_1 \sim 30 D_1, \quad A_0 - A_2 \sim 21D_1 + D_4, \quad  
A_0-A_3\sim 17D_1 + D_4,\\ 
B_1+B_2 - 2 A_0 \sim 29D_1 + D_4,
\end{gather*}
then the points $X_1(n_1,n_2,n_3)$ and $X_2(n_1,n_2,n_3)$
where the wave 
function $\psi(n_1,n_2,n_3)$ 
has additional zeros can be found from Table~\ref{tab:jakobianN} and 
from equation
\begin{equation} \label{eq:X1X2}  X_1(n_1,n_2,n_3) + X_2(n_1,n_2,n_3) - 2 \infty 
 = [n D_1 + m D_4]_r ,
\end{equation}
where $n\in\{ 0,1,\dots,30 \}$ and $ m\in\{ 0,1 \}$ are given by 
\begin{align} \label{eq:mn-n1}
&n  =  29 +30 n_1+ 21 n_2+17 n_3)  \mod 31, \\ 
&m  =  1 + n_2+n_3  \mod 2. \label{eq:mn-n2}
\end{align}
\begin{Rem}
Notice that the infinity point $\infty$ is the Weierstrass point of the curve
$\calC$, which violates the assumption of general position of points used 
in the
construction of solutions of the discrete KP equation. This will not change
the Jacobian picture of the construction but in some situations, which we will
point out, will affect uniqueness of the wave function. We remark that such a
choice is indispensable in reduction of the method from the discrete KP 
equation to the discrete KdV equation (see, for example \cite{KWZ,BialDol-d-KP-KdV}).
\end{Rem}
To find effectively the wave function we will constraint the range of 
parameters from
$\ZZ^3$ to the parameters of the group of $\FF_7$-rational points of the
Jacobian. Let us introduce functions $h_1$ and $h_4$ corresponding to generators
of the two cyclic subgroups of $J(\calC;\FF_7)$. The function $h_1$ with the
divisor $31 D_1$ and normalized at the infinity point is equal to
the product
$\prod_{i=0}^{30} g_0(i)$ 
and reads
\begin{multline*}
h_{1} =
1+2 u+{u^2}+4 {u^3}+3 {u^5}+{u^6}+ 3 {u^7}+ {u^8}+ 4 {u^9}+\\ 4 {u^{10}}+
2 {u^{11}}+5 {u^{12}}+2 {u^{13}}+ 4 {u^{14}}+
3 {u^{15}}+\big(5 u+2 {u^2}+ \\ 5 {u^3}+ 4 {u^5}+6 {u^6}+4 {u^7}+3
{u^9}+5 {u^{10}}+5 {u^{11}}+4 {u^{12}}+{u^{13}}\big) v ,
\end{multline*}
where we also used equation of the curve \eqref{eq:curve} to reduce higher order
terms in $v$.
The normalized function $h_4$ with the divisor $2D_4$ is 
\[ h_4 = u-6.
\]
Let us introduce other auxiliary functions $f_2$ and $f_3$
to factorize the zeros of the wave
function at $A_2$ and $A_3$.
Notice that 
\[ A_2 + 21 D_1 + D_4 - \infty \sim 0,
\]
which implies that there exists a polynomial function on 
$\calC$ with simple zero at $A_2$ and other zeros in the distinguished (by our
choice of description of $J(\calC;\FF_7)$) points $(1,1)$ and $(6,4)$.
Define $f_2$ as the unique such function normalized at the infinity point
$\infty$, then
\begin{multline*}
f_2 =1+5 u+{u^2}+4 {u^4}+6 {u^5}+4 {u^6}+4 {u^7}+3 {u^8}+4 {u^9}+ \\6 {u^{11}}+
\big(6+4 u+2 {u^2}+5 {u^3}+6 {u^4}+6 {u^6}+{u^7}+{u^8}+{u^9}\big) v.
\end{multline*}
Similarly we define the normalized function
\begin{equation*}
f_3 = 1+6 u+2 {u^2}+6 {u^5}+{u^6}+5 {u^7}+5 {u^8}+4 {u^9}+ 
\big(4+3 u+5 {u^2}+4{u^5}+2 {u^6}+{u^7}\big) v,
\end{equation*}
with the divisor $A_3+17D_1+D_4 -\infty$.

Uniqueness of the wave function $\psi$ implies that it
can be decomposed as follows
\begin{equation} \label{eqn:psi-W}
 \psi(n_1,n_2,n_3)=\frac{f_2^{n_2} f_3^{n_3}}{ h_{1}^p h_4^q} W(m_1,m_2),
\end{equation}
where $W(m_1,m_2)$ is the unique normalized function with the divisor
\begin{equation} \label{eq:X1X2-01}
m_1 D_1+ m_2 D_4 +Y_1(m_1,m_2) + Y_2(m_1,m_2) - 
(12,6) - (47,6),
\end{equation}
where
\begin{equation}
Y_1(m_1,m_2) + Y_2(m_1,m_2) = X_1(n_1,n_2,n_3) + X_2(n_1,n_2,n_3),
\end{equation}
and the new variables  $m_1$ i $m_2$ are given by 
\begin{eqnarray} 
\label{eq:mi-n2}
 21 n_2 + 17 n_3 - n_1 &=& 31p-m_1, \quad m_1\in\{ 0,1,\dots,30 \}, \\
\label{eq:mi-n1}
 n_2+n_3 &=& 2q-m_2, \quad m_2\in\{ 0,1 \}. 
\end{eqnarray} 

To find  the functions $W(m_1,m_2)$ for all $m_1\in\{ 0,1,\dots,30 \}$ 
and $ m_2\in\{ 0,1 \}$ let us first notice that $W(0,0)= 1$ and $W(0,1)$ is 
indeed the
function found in equation \eqref{eq:W01-2}. 
For $m_1\in\{ 1,\dots,30 \}$ 
and $ m_2\in\{ 0,1 \}$  define the functions $w_{m_2}(m_1)$ as follows
\begin{equation*}
    W(m_1,m_2) = w_{m_2}(m_1) W(m_1-1,m_2).
\end{equation*}
Equations \eqref{eq:g}, \eqref{eq:X1X2}-\eqref{eq:mn-n2} and
\eqref{eq:X1X2-01}-\eqref{eq:mi-n1} imply that for such range of $m_1$ and $m_2$
we have 
\begin{equation*}
g_m(n) = w_{m_2}(m_1), 
\end{equation*}
where
\begin{equation*}
m_2 = 1 - m \mod{2} , \quad m_1 = 29 - n \mod{31}.
\end{equation*}
Finally, under identification $w_{m_2}(0)=W(0,m_2)$ we obtain
\begin{equation*}
	W(m_1,m_2)=\prod_{i=0}^{m_1} w_{m_2}(i) , 
\end{equation*}
which, together with factorization \eqref{eqn:psi-W}, gives the wave function
$\psi$ for all values of $(n_1,n_2,n_3)\in \ZZ^3$. 

\begin{Rem}
For $(m_1,m_2)=(29,1)$ we have $X_1=X_2=\infty$. Because the infinity point
$\infty$ is the Weierstrass point of order two, there exist functions with
divisor of poles equal to $2\infty$. This means that $\psi$ is not uniqely 
determined in this case. However it is natural to keep the divisor of $\psi$,
and therefore $\psi$ itself, exactly like it is given from the flow on Jacobian.
Notice that because for $X_1=X_2=\infty$ we stay in the divisor $\WW_{g-1}$ 
then this ambiguity does not affect construction of the $\tau$-function.   
\end{Rem}

\begin{figure}
\begin{center}
\leavevmode\epsfysize=5.7cm\epsffile{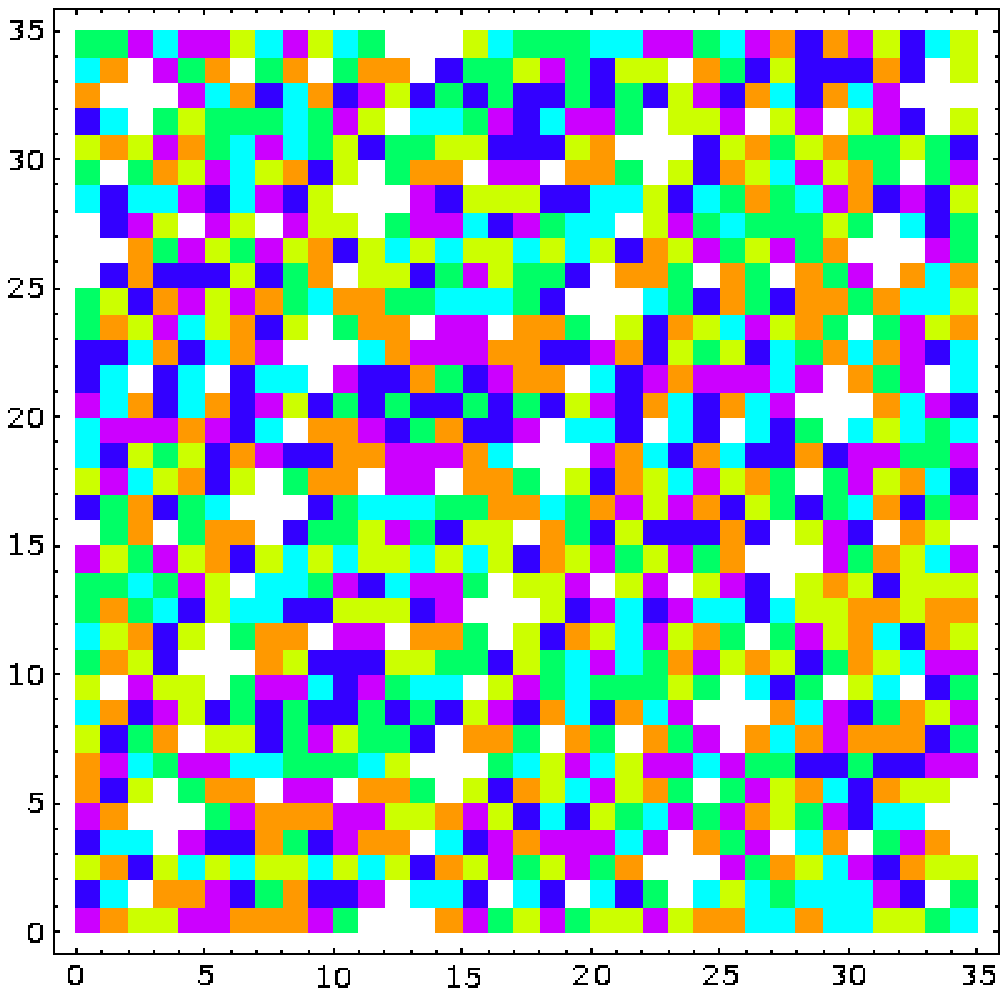} \hspace{0.1cm}
\leavevmode\epsfysize=5.7cm\epsffile{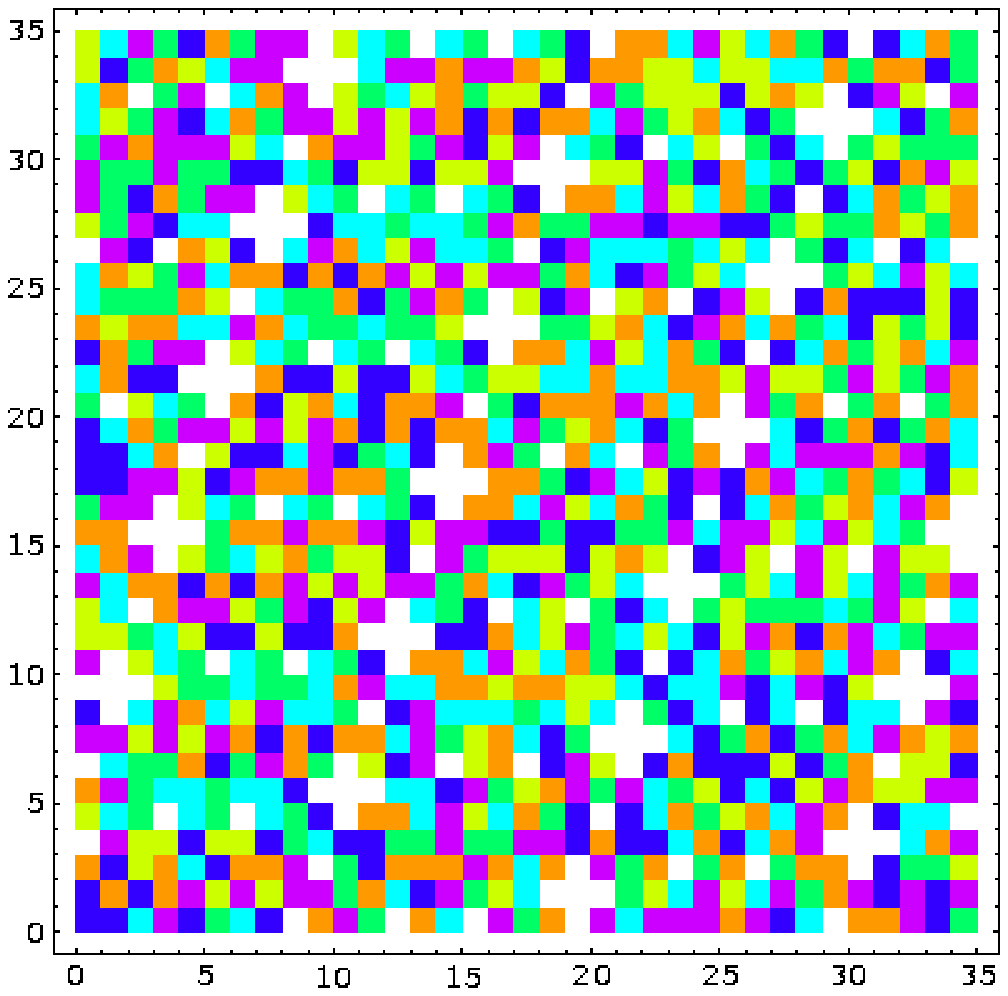}
\bigskip
\leavevmode\epsfysize=5.7cm\epsffile{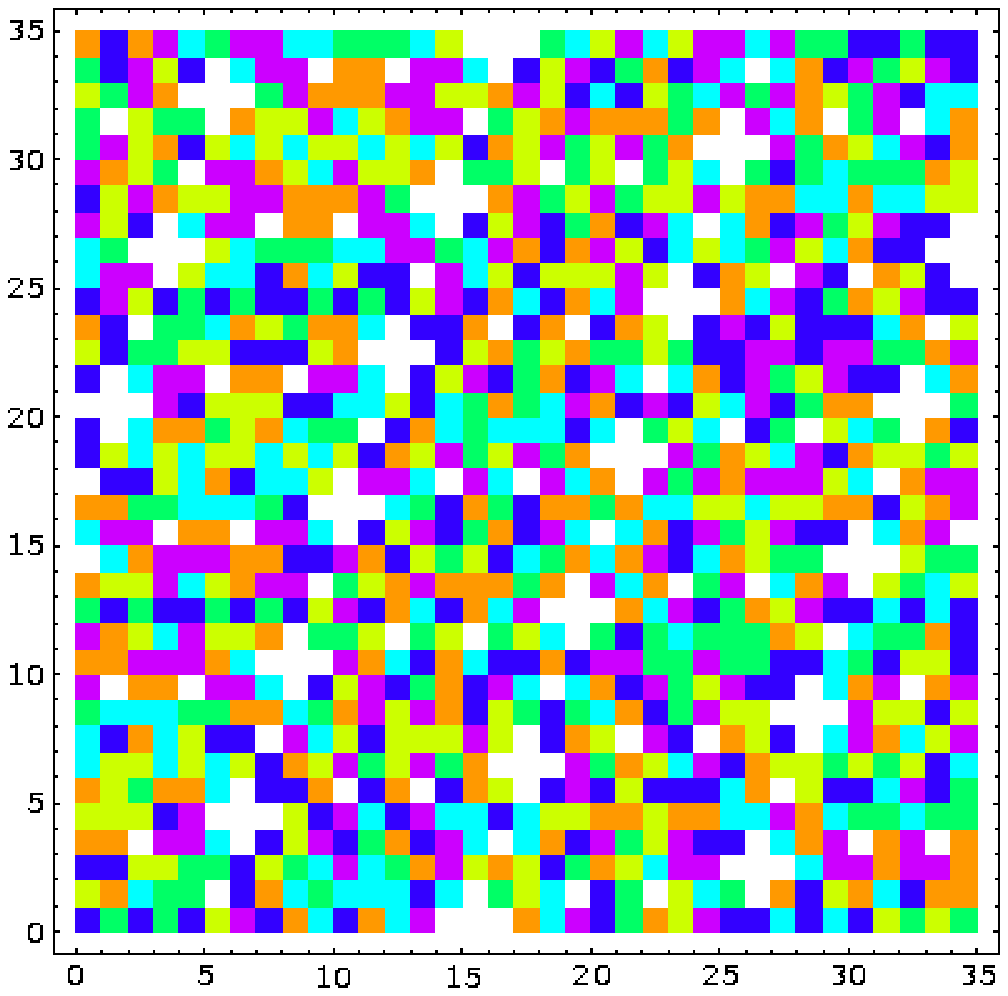} \hspace{0.1cm}
\leavevmode\epsfysize=5.7cm\epsffile{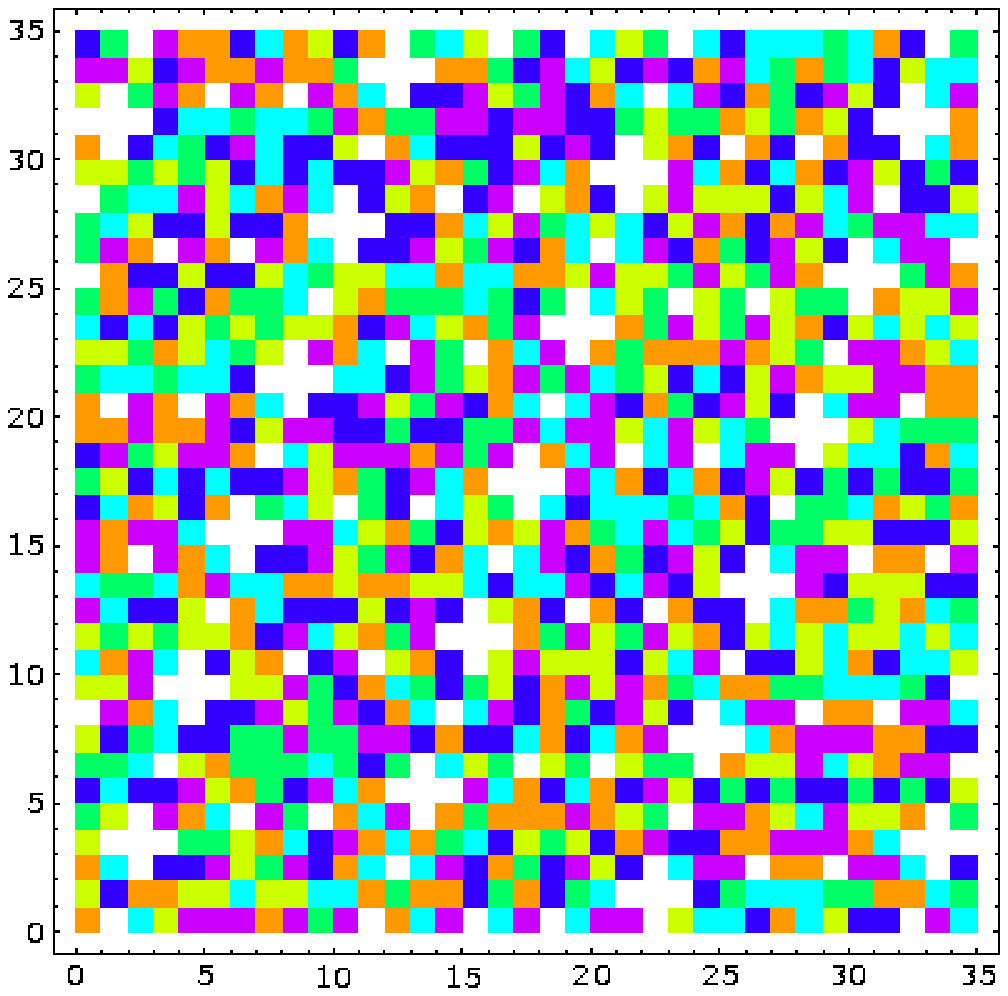} 
\mbox{
\epsfysize=0.35cm\epsffile{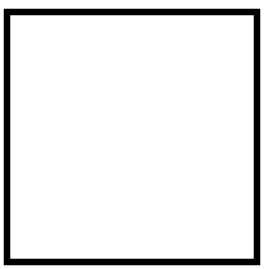} -- $0$, 
\epsfysize=0.35cm\epsffile{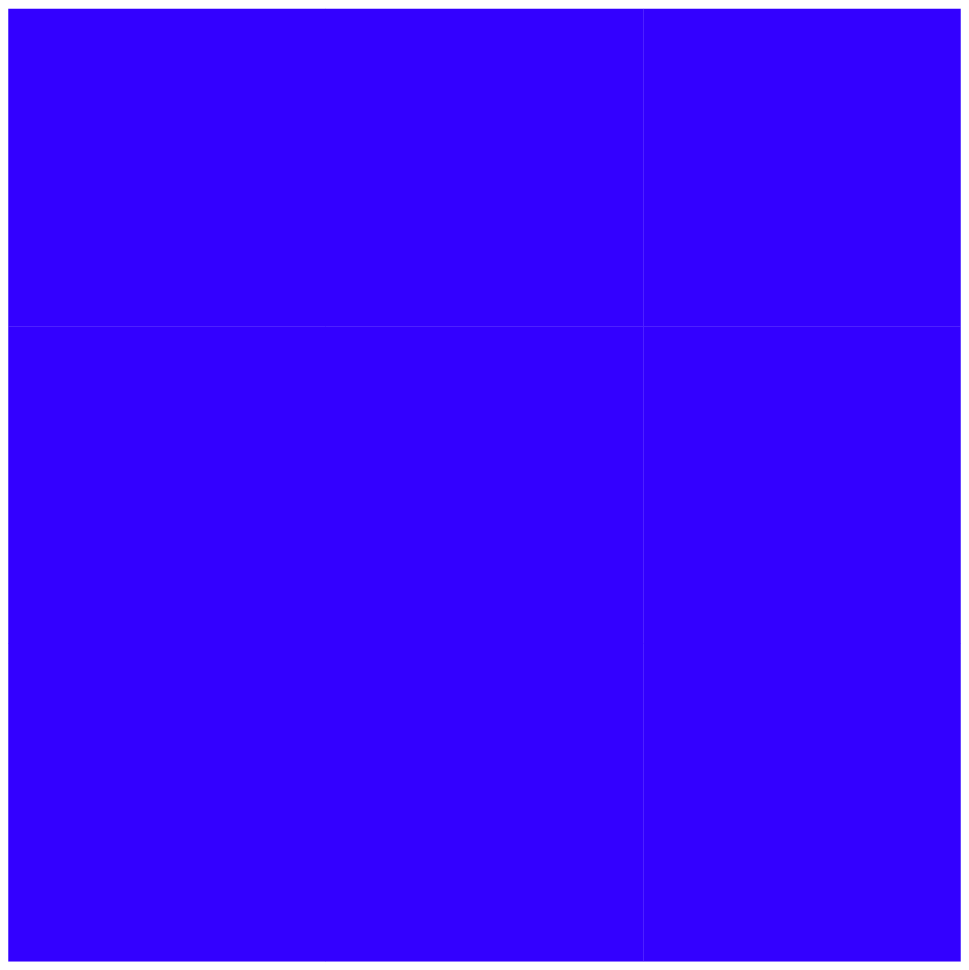} -- $1$, 
\epsfysize=0.35cm\epsffile{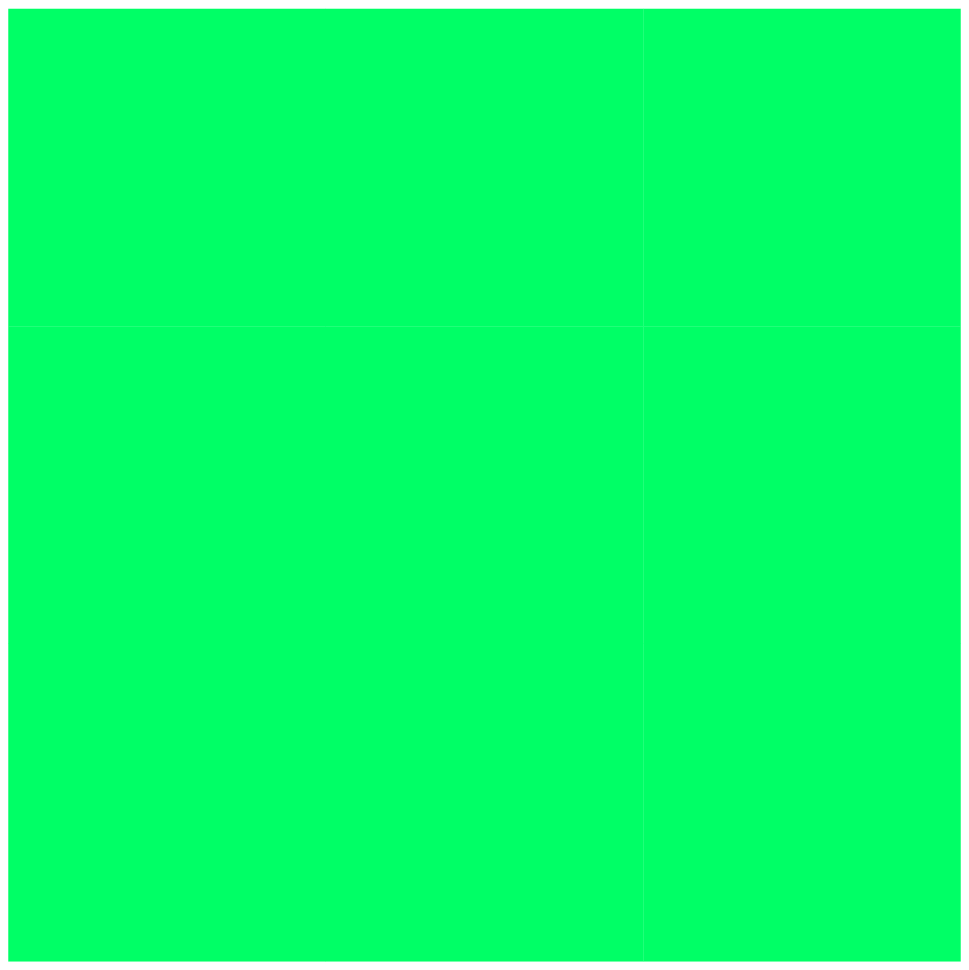} -- $2$,
\epsfysize=0.35cm\epsffile{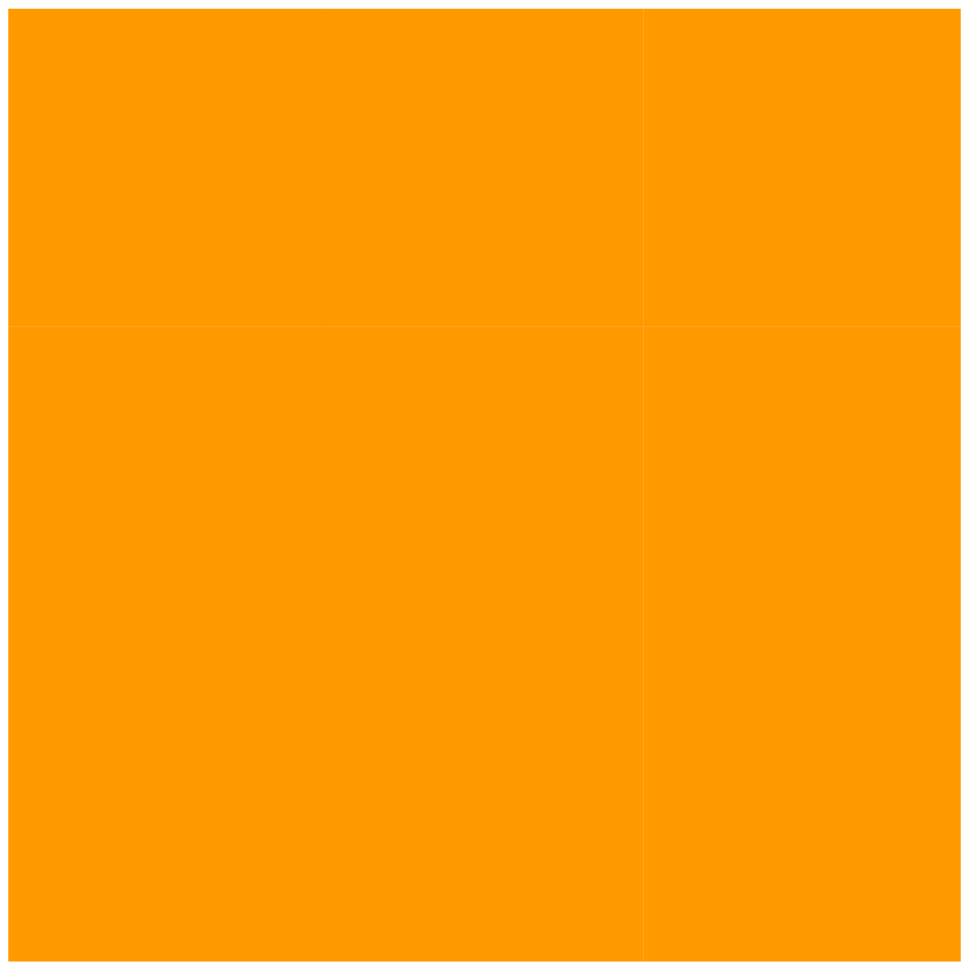} -- $3$, 
\epsfysize=0.35cm\epsffile{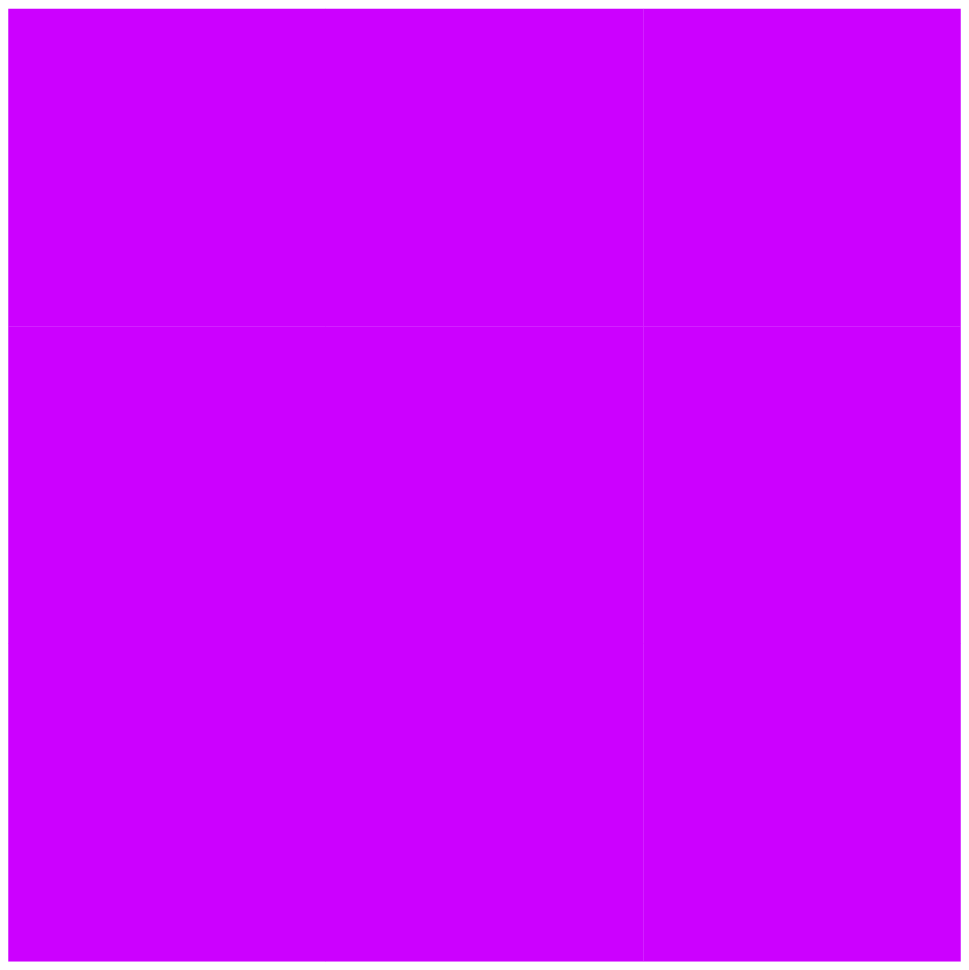} -- $4$, 
\epsfysize=0.35cm\epsffile{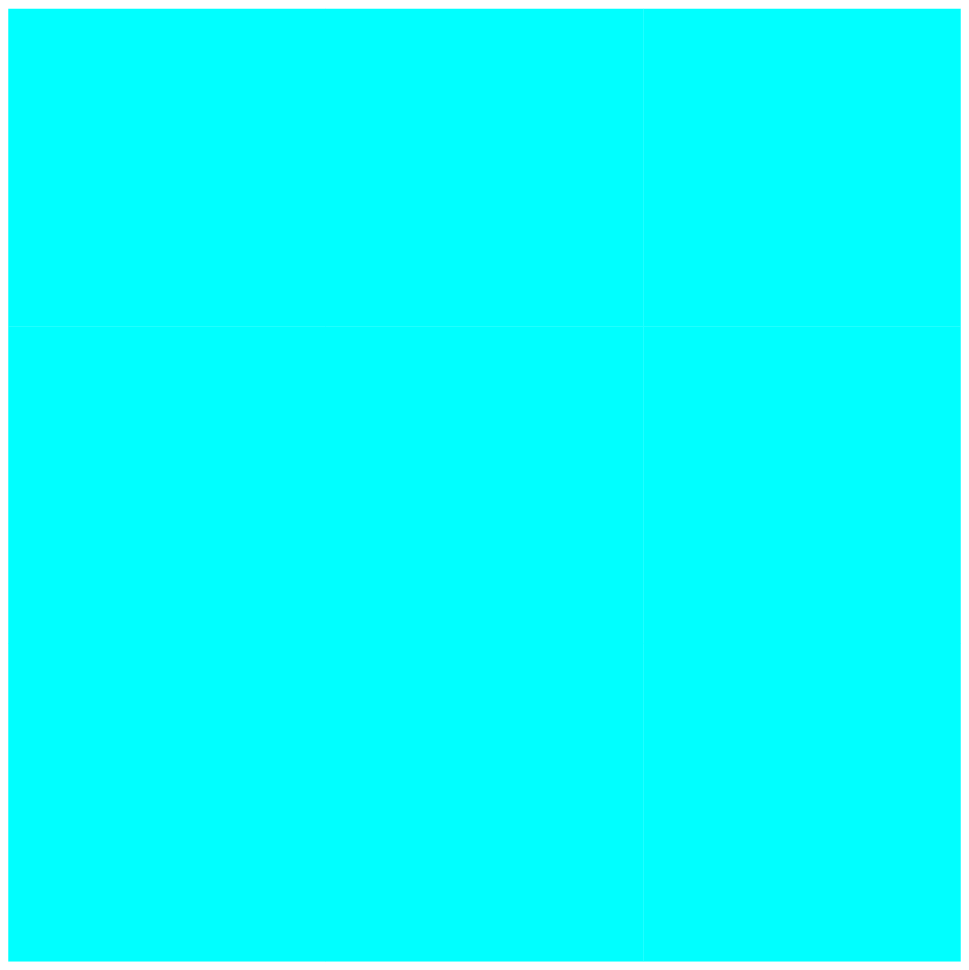} -- $5$, 
\epsfysize=0.35cm\epsffile{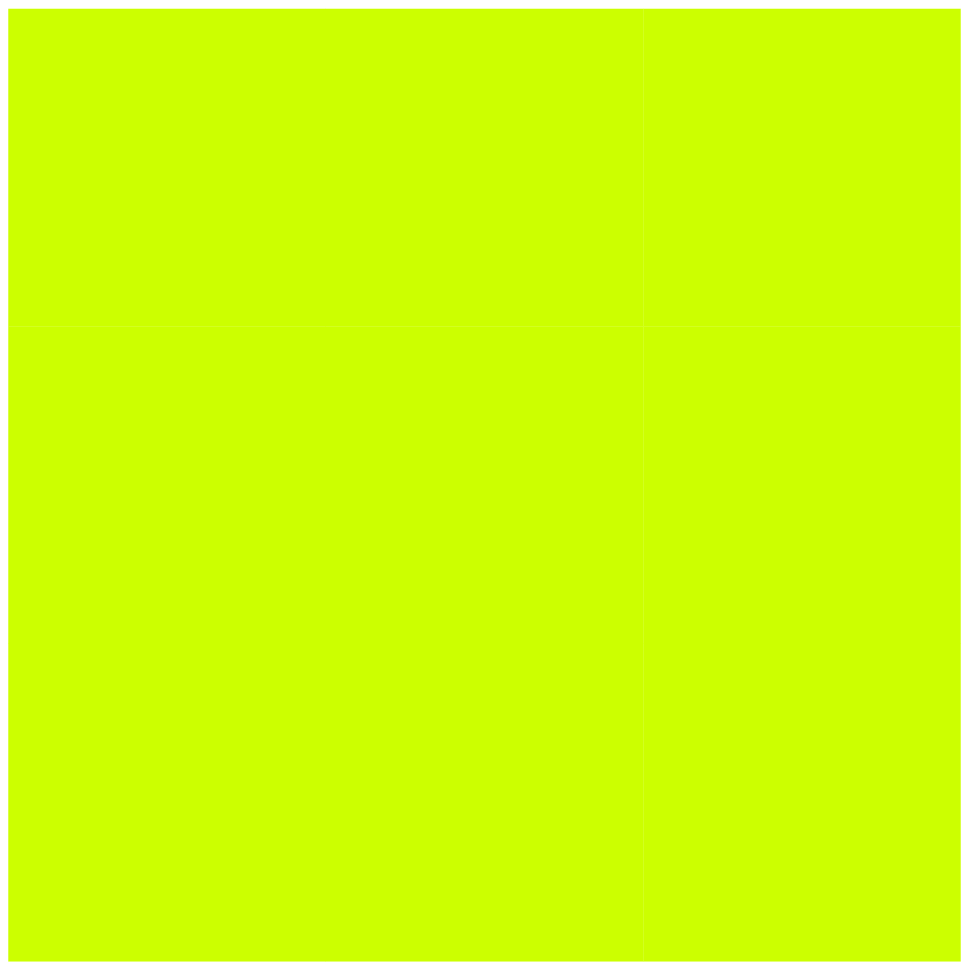} -- $6$.
} 
\end{center}
\caption{$\FF_{7}$-valued solutions of discrete KP equation out 
of genus two hiperelliptic curve $\calC$. 
Variables $n_1$ (directed to the right) and $n_2$ (directed up) 
take values from $0$ to $34$.
Subsequent figures are for values of $n_3=-1,0,1,2$. 
}
\label{fig:KPg2}
\end{figure}

The coefficients ${\zeta_0}^{(k)}(n_1,n_2,n_3)$, $k=1,2,3$, of expansion of the
wave function can be obtained from factorization \eqref{eqn:psi-W}
and are given by
\begin{eqnarray}
\label{eqn:z-z1}
 \zeta_0^{(1)} (n_1,n_2,n_3)&=& 
 6^{n_3+p} 4^q \left(\frac{W(m_1,m_2)}{t_1^{m_1}}\right)\Bigr\rvert_{t_1=0} ,\\
\label{eqn:z-z2}
 \zeta_0^{(2)} (n_1,n_2,n_3)&=& 
 6^{n_2} 5^{n_3+q} 4^p \: W(m_1,m_2)|_{t_2=0} ,\\
\label{eqn:z-z3}
 \zeta_0^{(3)} (n_1,n_2,n_3)&=& 
 5^{n_2+n_3} 3^p 6^q \: W(m_1,m_2)|_{t_3=0}. 
\end{eqnarray}
Using definition of the $\tau$ function for nonzero $\rho_i$, i.e. 
equation~\eqref{eq:tau-def},
and putting $\tau=0$ for points of the divisor $\WW_{g-1}$
we obtain the corresponding $\FF_7$-valued solution of the discrete KP
equation~\eqref{eq:tauKP}. 
This $\tau$-function is presented in Figure~\ref{fig:KPg2}. 
Notice that due to quasi-periodicity of the $\tau$-function, we have 
to calculate the solution in this way only for a finite range of values of 
the variables.

The periods $\Pi_i$, $i=1,2,3$, of zeros of the $\tau$-function are,
respectively, $31$, $62$ and $62$.  
Equivalently, the "period vectors" of zeros can be choosen as
\begin{equation*}
v_1=\left( \begin{array}{c} -4 \\ -1 \\  1 \end{array} \right) \quad
v_2=\left( \begin{array}{c} 11 \\ 2 \\  0 \end{array} \right) \quad
v_3=\left( \begin{array}{c} 2 \\ 6 \\  0 \end{array} \right). 
\end{equation*} 
Because $c_{(1)1} = 6$, $c_{(2)1} = 3$ and $c_{(3)1} = 5$ then periods of the
functions $\rho_i$, $i=1,2,3$, in that variable are, respectively, $2\cdot 31$,
$3\cdot 31 \cdot 2$ and $6\cdot 31$. Moreover, we have $\tau(31,0,0)=3$ which gives 
$\tau(62,0,0)=5=3^2\cdot 6^{31}\mod{7}$, and therefore the period of $\tau$ in $n_1$ is $6\cdot
2\cdot 31$.

\section*{Acknowledgments}
The paper was partially supported by the 
University of Warmia and Mazury in Olsztyn under the grant  522-1307-0201
and by KBN grant 2~P03B~12622.

\bibliographystyle{amsplain}

\providecommand{\bysame}{\leavevmode\hbox to3em{\hrulefill}\thinspace}
\providecommand{\MR}{\relax\ifhmode\unskip\space\fi MR }
\providecommand{\MRhref}[2]{%
  \href{http://www.ams.org/mathscinet-getitem?mr=#1}{#2}
}
\providecommand{\href}[2]{#2}

\end{document}